\documentclass[referee]{raa}      
\usepackage{graphicx,times}
\usepackage{natbib}
\usepackage{amssymb,amsmath}
\usepackage[figuresright]{rotating}
\usepackage{rotating} 
\usepackage{color}
\usepackage{multirow}
\usepackage{longtable,booktabs}
\usepackage[titletoc]{appendix}
\usepackage{adjustbox}
\usepackage{rotfloat}
\usepackage{threeparttable}
\usepackage{times}
\usepackage{lscape}
\usepackage[margin=2cm]{geometry}
\bibpunct{(}{)}{;}{a}{}{,}

\usepackage[a4paper=true,driverfallback=dvipdfm,pagebackref=true]{hyperref}
\hypersetup{pdftitle = The title of my PDF, pdfauthor = My name, pdfsubject= The subject, pdfkeywords = keyword1 keyword2 keyword3} 
\hypersetup{colorlinks = true, linkcolor = green, anchorcolor = red, citecolor = blue, filecolor = red, pagecolor = red, urlcolor = red}

\begin{document}

   \title{Chemical abundances of three new Ba stars from the Keck/HIRES spectra}

 \volnopage{ {\bf 20XX} Vol.\ {\bf X} No. {\bf XX}, 000--000}
   \setcounter{page}{1}

   \author{Shuai Liu\inst{1,2}, Liang Wang\inst{3,4}, Jian-Rong Shi\inst{1,2},  Zhen-Yu Wu\inst{1,2},  Hong-Liang Yan\inst{1,2}, Qi Gao\inst{1,2}, Chun-Qian Li\inst{1,2}
   }

\institute{CAS Key Laboratory of Optical Astronomy, National Astronomical Observatories, Chinese Academy of Sciences, Beijing 100101, China; {\it sjr@nao.cas.cn}\\
 \and     School of Astronomy and Space Science, University of Chinese Academy of Sciences, Beijing 100049, China\\ 
\and    National Astronomical Observatories / Nanjing Institute of Astronomical Optics \& Technology,Chinese Academy of Sciences, Nanjing 210042, China\\
\and    CAS Key Laboratory of Astronomical Optics \& Technology, Nanjing Institute of Astronomical Optics \& Technology, Chinese Academy of Sciences, Nanjing 210042, China\\
\vs \no
  {\small Received 20XX Month Day; accepted 20XX Month Day}
}

\abstract{Based on high resolution, high signal-to-noise (S/N) ratio spectra from Keck/HIRES, we have determined  abundances of 20 elements for 18 Ba candidates. The parameter space of these stars are in the range of 4880 $\leq$ $\rm{T_{eff}}$ $\leq$ 6050\,K, 2.56 $\leq$ log $g$ $\leq$ 4.53\,dex and $-$0.27 $\leq$ [Fe/H] $\leq$ 0.09\,dex. It is found that four of them can be identified as Ba stars with [s/Fe] $>$ 0.25\,dex (s: Sr, Y, Zr, Ba, La, Ce and Nd), and three of them are newly discovered, which includes two Ba giants (HD\,16178 and HD\,22233) and one Ba subgiant (HD\,2946). Our results show that the abundances of $\alpha$, odd and iron-peak elements (O, Na, Mg, Al, Si, Ca, Sc, Ti, Mn, Ni and Cu) for our program stars are similar to those of the thin disk, while the distribution of [hs/ls] (hs: Ba, La, Ce and Nd, ls: Sr, Y and Zr) ratios of our Ba stars is similar to those of the known Ba objects. 
None of the four Ba stars show clear enhancement in carbon including the known CH subgiant HD\,4395. 
It is found that three of the Ba stars present clear evidences of hosting stellar or sub-stellar companions from the radial velocity data. 
\keywords{Galaxy:Ba stars:Chemical abundances:Binary}
} 
\authorrunning{Shuai Liu et al. }            
\titlerunning{Chemical abundances of three new Ba stars}  
\maketitle

%
\section{Introduction}           
\label{sect:intro}
Barium stars are a type of particular objects, which were first discovered by \cite{1951ApJ...114..473B}, and  exhibit strong spectral lines from slow neutron capture process (s-process) elements (such as Sr, Y, Zr, La, Ce, Pr, Nd or Sm.) and CH, CN and $\rm{C_2}$ molecules. Early on, it was noted that Ba stars are G/K-type giants. Later, similar spectral features have been found in some CH subgiant \citep{1974ApJ...194...95B} and dwarf stars \citep{1994A&A...281..775N}, these objects are called Ba dwarfs.  
It is well known that the s-process nucleosynthesis happens in the thermal pulse Asymptotic Giant Branch (TP-AGB) \citep[e.g.][]{1999ARA&A..37..239B,2005ARA&A..43..435H,2011RvMP...83..157K,2014PASA...31...30K} stage, in which the inner materials are carried to the surface by the Third Dredge-Up (TDU) mechanism. However, the observed luminosities of Ba stars failed to attain the threshold to trigger this stage \citep{2017A&A...608A.100E,1997A&A...321L...9B}. Thus, this type stars have enhanced their heavy elements by the mechanism of self-enrichment. 

\cite{1980ApJ...238L..35M} discovered that the radial velocities of their nine of eleven strong Ba stars present periodic variations, and suggested that those objects may have a companion. Subsequently, observations on the variances of radial velocities \citep[e.g.][]{1983ApJ...268..264M,1990ApJ...352..709M,1998A&A...332..877J} and UV excess \citep[e.g.][]{1984ApJ...278..726B,2000ApJ...533..969B,2011AJ....141..160G} for some Ba stars confirmed this suggestion, and indicated that the companion star is quite possible a white dwarf. Therefore, a deduction could be drawn that Ba stars formed by the accretion of heavy elements from a companion, which have already undergone the TP-AGB stage \citep[e.g.][]{1980ApJ...238L..35M,1995MNRAS.277.1443H}.

Recently, some works noted that the degrees of Ba enrichment may be influenced by the distance between the two stars \citep{1995MNRAS.277.1443H,2004ARep...48..597A,2016RAA....16...19Y}, or the metallicities \citep{1998A&A...332..877J}, the later plays a critical role on the s-process nucleosynthesis of secend-to-first peak elements in AGB stars \citep{1998ApJ...497..388G,2000A&A...362..599G}. According to the evolutionary timescales and the orbital characteristics of Ba stars \citep{2000IAUS..177..269N} and the cooling times of white dwarfs \citep{2000ApJ...533..969B}, it can be inferred that most Ba stars are contaminated while in the main sequence, which means that the progenitor star of a Ba giant might be a Ba dwarf. However, the mass distribution of the observed Ba dwarfs (M $<$ 2$\rm{M_{\odot}}$) peaks at much lower mass compared to those of Ba giants (M $>$ 2$\rm{M_{\odot}}$), and Ba dwarfs have a distribution tending to more metal-poor than that of Ba giants ($-$0.1 $<$ [Fe/H] $<$ 0.1), which means that the Ba dwarfs we have observed do not represent the progenitor of the Ba giants \citep{2019A&A...626A.128E,2019A&A...626A.127J}. Some researches have found that Ba stars can also exist in triple systems, such as, HD\,48565 and HD\,114520 \citep{2000IAUS..177..269N,2019A&A...626A.128E}.

Heavy elements beyond Fe group  can also be produced by fast neutron capture process (r-process), however, the stellar site of main r-process  is still in debate. The candidate sites includes the neutrino-driven wind of core collapse supernovae \citep[CCSNe,][]{2013ApJ...770L..22W}, the magneto-hydrodynamically driven jet from rapidly rotating, strongly magnetized CCSNe \citep{2012ApJ...750L..22W}, and neutron star mergers \cite[e.g.][]{2004A&A...416..997A}. In some metal-poor stars, the abundance pattern shows an obvious enrichment of heavy elements between the s- and r-processes \citep{1997A&A...317L..63B,2006A&A...451..651J}, which is called i-process \citep{1977ApJ...212..149C}.  Therefore, Ba stars perhaps have more complicated mechanism of enrichment.

Chemical abundance is a invaluable tool to probe the nucleosynthetic process, and the [hs/ls] ratio has been widely used to investigate the efficiency of the s-process \citep{1991ApJS...77..515L,2007A&A...468..679S}.  Here `hs' represents mean value of the heavy s-process element abundances (Ba, La, Ce and Nd), while `ls' is for light elements (Sr, Y, Zr). The theoretical models arrived at, after AGB stage, the ratios of ls, hs and [hs/ls] decrease with increasing metallicity during $-$1.5 $<$ [Fe/H] $<$ 0\,dex for stars with mass between 1.5 $\rm{M_{\odot}}$ and 3 $\rm{M_{\odot}}$ \citep{2021ApJ...908...55B}. \cite{2007A&A...468..679S} found that the strong Ba stars have higher [hs/ls] than those of the mild Ba stars, and \citet{2016MNRAS.459.4299D} showed that the [s/Fe] and [hs/ls] ratios are strongly anticorrelated with the metallicity \citep[also see][]{2018MNRAS.474.2129K}. It is suggested by \cite{2020JApA...41...37S} that the ratio of [Rb/Zr] is an important diagnostic index for understanding the average neutron density in the s-process site and the mass of the AGB stars, the [Rb/Zr] ratio is negative for low-mass AGB stars, while it is positive for intermediate-mass AGB stars  \citep[e.g.][]{2012A&A...540A..44V}.

Based on the strength of the Ba\,II resonance line at 4554\,\AA\,, \cite{1965MNRAS.129..263W} classified the Ba stars from the scale 0 (weakest) to 5 (strongest), and \cite{1991AJ....101.2229L} noted that the strong Ba stars of scale = 2$-$5 belong to the old disk with a large dispersion in space velocities,  while the weak ones of scale $<$ 2 are  generally young disk objects with a small scatter in space velocities. Although, \cite{2016MNRAS.459.4299D} noted that the Ba stars could not be sorted through the [s/Fe] ratios clearly,  he suggested that a value of [s/Fe] = 0.25\,dex is the minimal ratio for a object identified as a Ba star. Comparing to the mild Ba stars, strong ones generally have lower metallicities \citep{1985A&A...150..232K}, and tend to have shorter periods \citep{2019A&A...626A.127J}. 

Up to now, only a small number of Ba stars have been certified especially for Ba dwarfs. The catalog of 389 Ba giants listed by \cite{1991AJ....101.2229L} including certain and candidate samples, and some of them have no heavy-element overabundances \citep{1996A&A...306..467J}. \citet{2016MNRAS.459.4299D} gathered 182 Ba giant candidates, and confirmed 169 of them with s-process elements enhanced. \citet{2018MNRAS.474.2129K} have collected 58 Ba CH subgiant and dwarf stars with abundance information including their three new Ba dwarfs, however, the number is still smaller compared to the Ba giants. Therefore, it is important to enlarge the sample for understanding the formation and evolution of Ba dwarfs.
 
In previous work, we have found 18 Ba candidate stars based on analyzing the high resolution, high signal-to-noise-ratio spectra from Keck/HIRES \citep{2020ApJ...896...64L}. 
In this paper, we have derived the detailed elemental abundances and radial velocities for the 18 Ba candidates by analyzing the HIRES high-resolution spectra. The outline of this paper is as follows.
In Section 2, we will briefly introduce the key information about the spectra and our program stars. The stellar atmospheric parameters of our program stars are presented in Section 3. In Section 4 we show the detail elemental abundances, and the results of the elemental abundances and radial velocities are discussed in Section 5. The conclusions are drawn in Section 6.

\section{The SPECTRA and SAMPLE STARS}

The High Resolution Echelle Spectrometer (HIRES) is a high resolution visible light slit spectrograph installed at the 10\,m Keck telescope in 1996 \citep{1994SPIE.2198..362V} , and it was used by the California Planet Search program team to obtain high resolution spectrum with a average resolution of R $\approx$ 60,000 and high precision radial velocities of $\sim$1-3\,$\rm{m\ s^{-1}}$  \citep[CPS;][]{2010ApJ...721.1467H}.  The precise radial velocities are obtained by installing a iodine molecule ($\rm{I_2}$) vapor cell with constant temperature (50.0 $\pm$ $\rm{0.1^{\circ}C}$) in front of the slit. The $\rm{I_2}$ imprints a series of weak absorption lines over the stellar spectrum, which can be used to precisely calibrate the wavelength \citep{1995PASP..107..966V,1996PASP..108..500B}. The wavelength coverages of the spectra are divided into three sections of 3640 to 4790\,\AA, 4970 to 6420\,\AA\ and 6540 to 7980\,\AA. All of the spectra in this study were taken from the HIRES archive, and the signal-to-noise (S/N) ratios for the spectra are listed in the last column of Table \ref{table1}. 

\begin{figure}[h]
\centering
\includegraphics[width=0.7\linewidth]{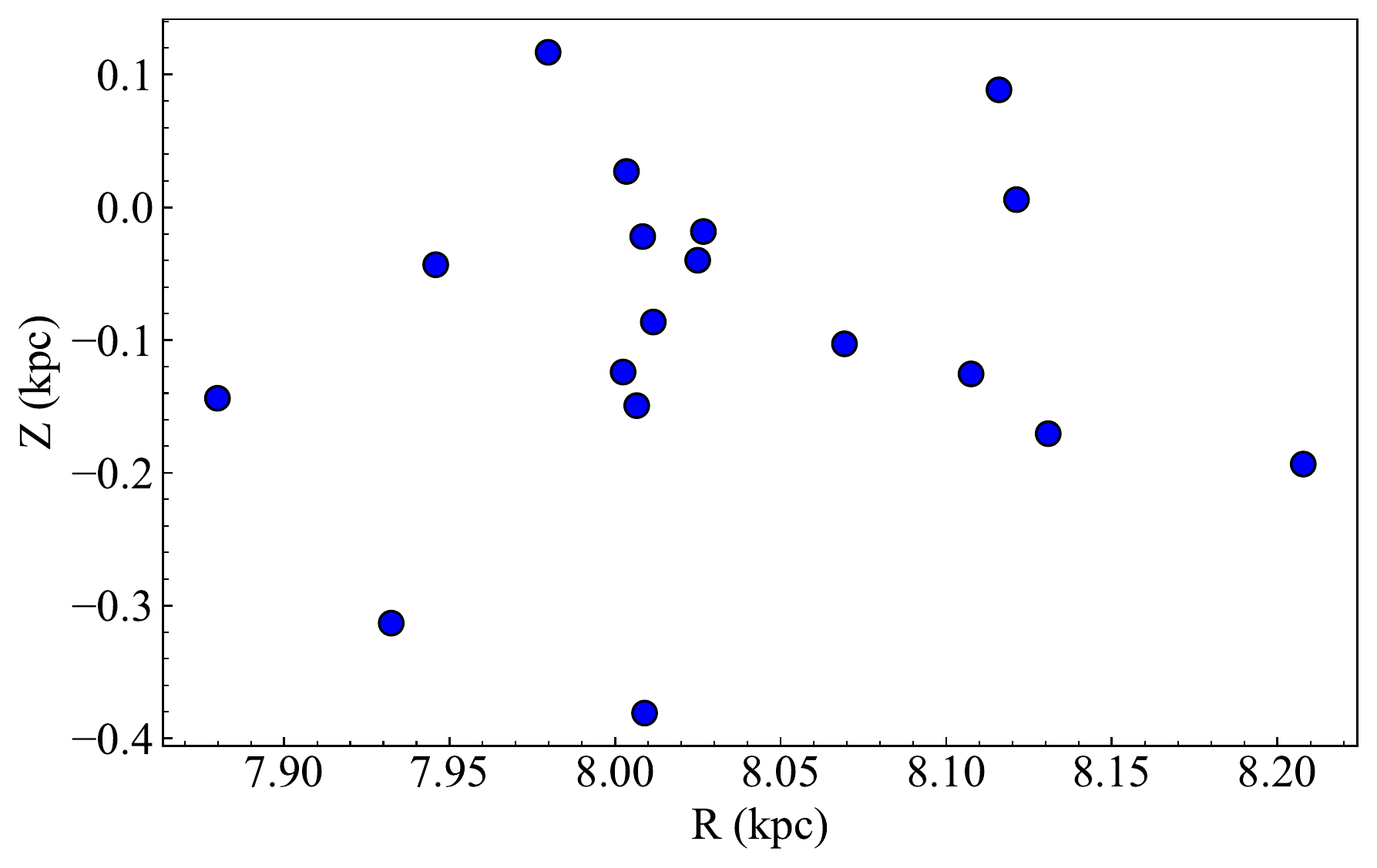}
\caption{The distribution of our sample stars in the R$-$Z plane.}
\label{figure1}
\end{figure}

The 18 Ba candidates have been selected from our previous work \citep{2020ApJ...896...64L}, which presented the Ba abundance of 602 objects derived from the spectra collected from CPS. Our sample includes the star HD\,4395, which was identified as a CH subgiant in previous studies \citep[e.g.][]{1999ARA&A..37..239B,2016MNRAS.459.4299D}.  Figure \ref{figure1} gives the spatial distribution in R $-$ Z plane for the program stars. Here, we use the 1/parallax as the distance of a star, and  the parallax comes from the Gaia EDR3 catalog \citep{2021A&A...649A...1G}. As shown in Figure \ref{figure1}, these stars lie in the solar vicinity, and most likely belong to the thin disk.

\begin{figure}
\centering
\includegraphics[width=1.0\linewidth]{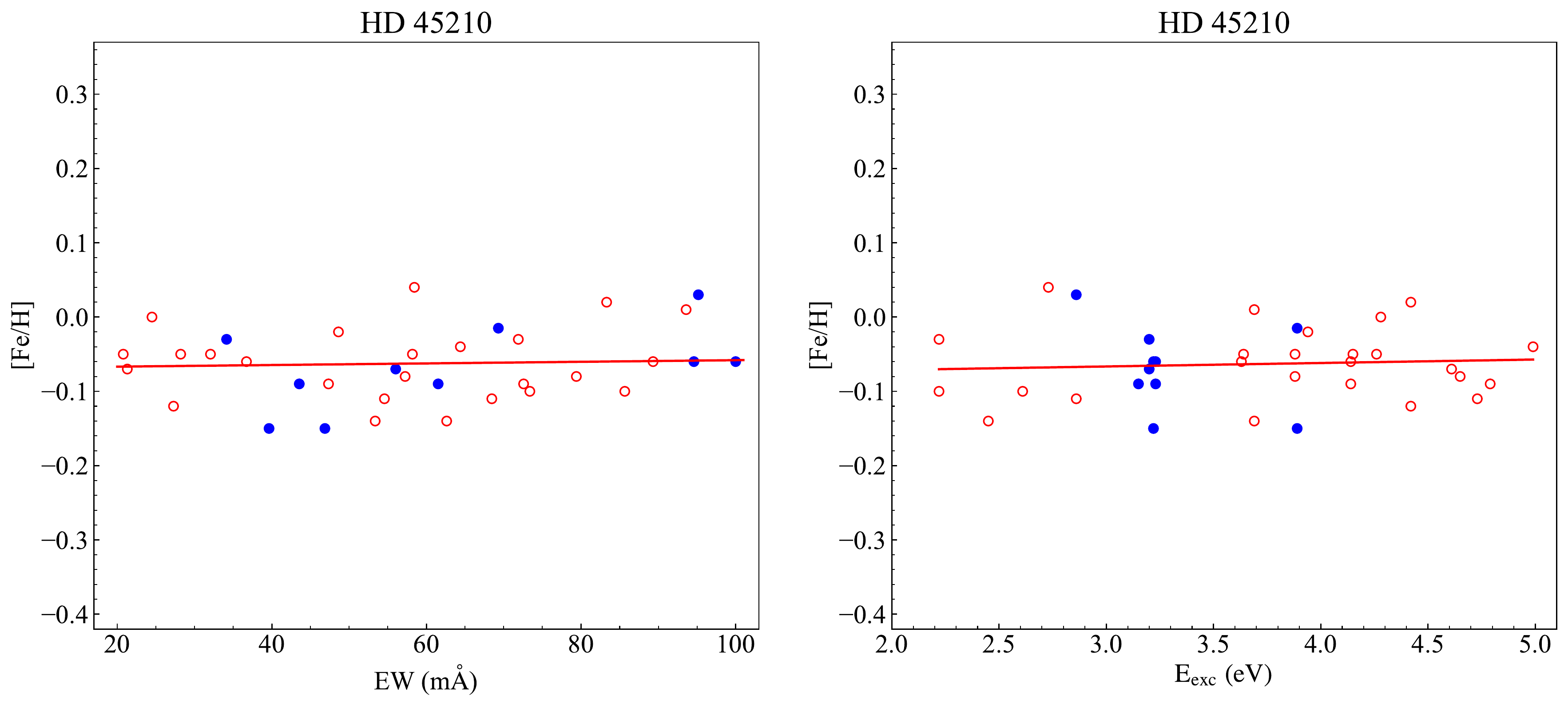}
\caption{The determination of stellar parameters of a typical star HD\,45210 based on the ionization and excitation equilibrium of Fe\,I and Fe\,II lines. Open red circles refer to Fe\,I lines and filled (blue) circles to Fe\,II lines. The red straight lines are the fitting results with a least square method.}
\label{figure2}
\end{figure}

\section{The Stellar atmosphere parameters}

In this paper, the stellar parameters of all 18 stars have been revised via the spectroscopic approach, e.g. the effective temperature is determined by fulfilling the excitation equilibrium of Fe\,I,  the surface gravity is obtained by the ionization equilibrium of Fe\,I and Fe\,II, and the microturbulence velocity is derived by forcing [Fe/H] from different Fe\,I lines to be independent of their equivalent widths. 
This process of deriving stellar parameters is an iterative procedure, and the initial data from \citet{2016ApJS..225...32B} is applied. In our analysis, 33 Fe\,I and 12 Fe\,II unblended optical lines selected from \cite{2018NatAs...2..790Y} have been adopted, and the information of these lines is shown in Table \ref{tableA}. 
It is noted that the departures from local thermal equilibrium (LTE) of Fe\,I lines are small, lower than 0.05\,dex for our sample stars \citep{2012MNRAS.427...50L, 2015ApJ...808..148S}, therefore, the non-local thermal equilibrium (NLTE) effects have not been considered when determining the iron abundance.  

Figure \ref{figure2} shows the derived abundances from individual Fe\,I and Fe\,II lines as functions of their equivalent widths (left panel) and excitation potentials (right panel) for a typical star of HD\,45210. The final stellar parameters of our 18 samples are presented in Table \ref{table1}. Based on multiple iterative processes, we estimate that the typical uncertainties of $T_{\rm eff}$,  $\rm{log\ g}$, [Fe/H], and $\xi_t$ are $\pm$80\,K, $\pm$0.1\,dex, $\pm$0.1\,dex, and 0.2\,km\,s$^{-1}$, respectively.

\begin{table}[htb]
\centering
\caption{ {\upshape The stellar parameters of our sample stars}}
\label{table1}
 \setlength{\tabcolsep}{2mm} 
\begin{tabular}{lrrrcc}
\hline
\hline
Star Name& $\rm{T_{eff}}$ & log g & [Fe/H] & $\xi_{t}$(km $\rm{s^{-1}}$) & S/N \\
\hline
HD\,2946  &   5570	&  3.70	&$-$0.27	  &1.11   &    80        \\
HD\,3458  &   5018	&  2.71 &$-$0.06	  &1.10   &    86      \\  
HD\,4395  &   5494	&  3.61 &$-$0.15	  &0.85   &    84     \\   
HD\,11131 &   5845	&  4.53 &0.00	      &0.70   &    130     \\  
HD\,12484 &   5835	&  4.50	&0.09	      &1.08   &    140       \\
HD\,16178 &   4900	&  2.72 &$-$0.09	  &1.15   &    99     \\   
HD\,18015 &   5603	&  3.64 &$-$0.11	  &1.05   &    90     \\   
HD\,18645 &   5458	&  3.36 &0.08	      &1.20   &    92     \\   
HD\,22233 &   5172	&  3.11 &$-$0.03	  &0.90   &    95       \\ 
HD\,38949 &   6050	&  4.50	&$-$0.04	  &1.00   &    93      \\  
HD\,45210 &   5717	&  3.54 &$-$0.07	  &1.10   &    89     \\   
HD\,72440 &   5600	&  3.60	&$-$0.08	  &1.00   &    86       \\ 
HD\,103847&   5209	&  4.52 &0.03	      &0.80   &    152     \\  
HD\,108189&   5273	&  3.20	&$-$0.09	  &0.95   &    93      \\  
HD\,200491&   5083	&  2.90	&0.03    	  &1.05   &    68       \\ 
HD\,205163&   4989	&  3.00	&0.09   	  &1.13   &    85      \\  
HD\,220122&   4880	&  2.56 &$-$0.18	  &1.08   &    93      \\  
HD\,224679&   5677	&  3.55 &0.08	      &1.05   &    86    \\   
\hline 
\end{tabular}
\end{table}


\begin{figure}
\centering
\includegraphics[width=0.7\linewidth]{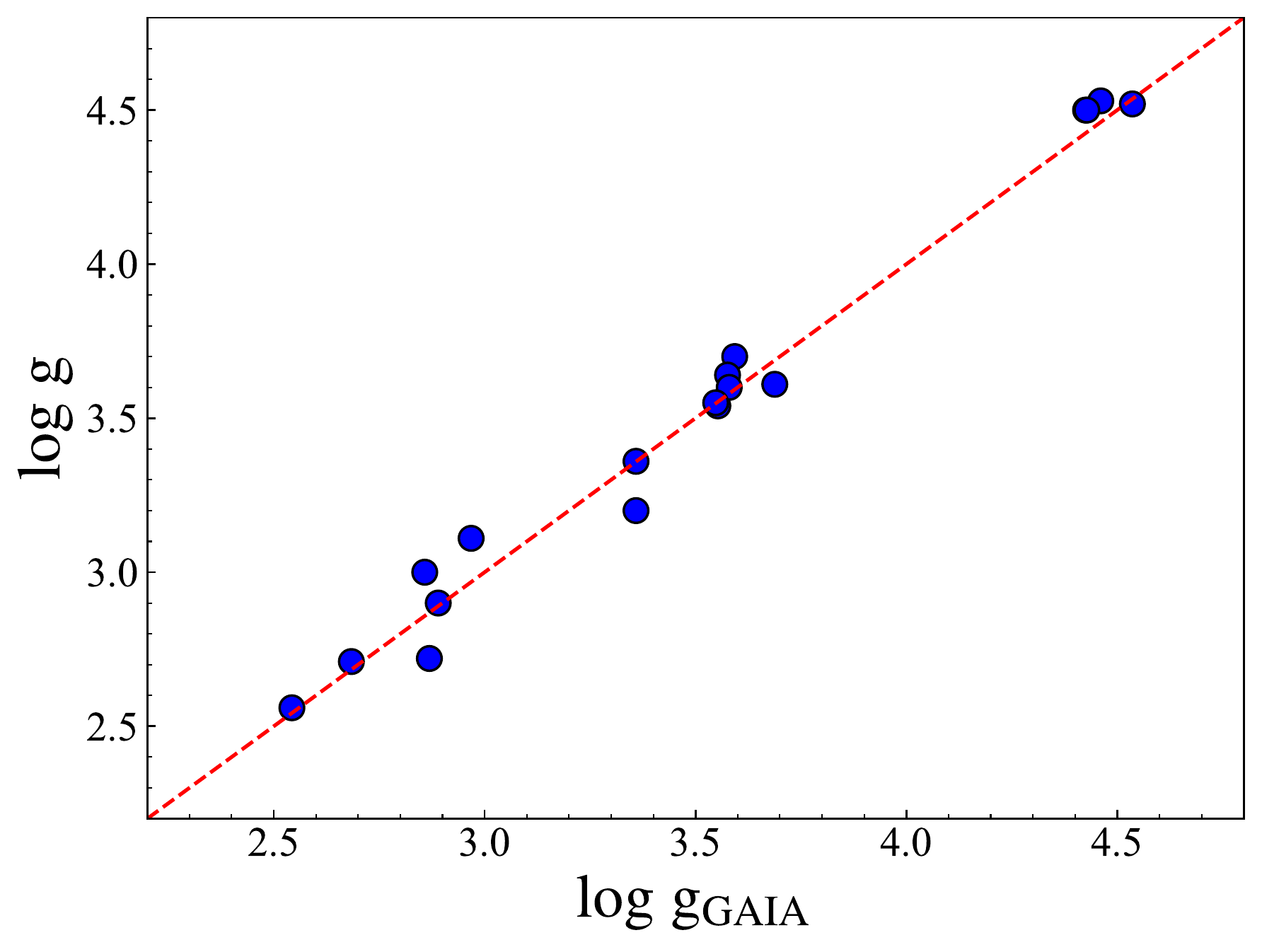}
\caption{The difference of the log g determined with our method and those calculated by PARAM. }
\label{figure3}
\end{figure}

We also determine the surface gravities (log\ g) adopted the parallax from the Gaia EDR3 \citep{2021A&A...649A...1G} using Girardi's online code PARAM 1.3$\footnote{\url{http://stev.oapd.inaf.it/cgi-bin/param_1.3}}$ \citep{2006A&A...458..609D}. 
As shown in Figure \ref{figure3}, there is a good agreement between these two values, and the mean scatter is about 0.07.  

\section{Analysis of element abundances}

\subsection{Atomic Data}
In our work, the line list along with the atomic data of C\,I, Na\,I, Mg\,I, Al\,I, Si\,I, Ca\,I, Sc\,II, Ti\,II, Cu\,I, Zr\,II, Sr\,II and Ba\,II are taken from \cite{2016ApJ...833..225Z}, and we adopted the solar abundances recommended by \cite{2009ARA&A..47..481A}. The oxygen abundance is derived by fitting the forbidden [O\,I] line at 6300\,\AA, which blended with a Ni\,I line, and the $\rm{log\,gf}$ values of the forbidden [O\,I] line and Ni\,I line at 6300\,\AA\ are adopted from \cite{2008ApJ...682L..61C}.
While, the line list of Mn\,I, Ni\,I, Sr\,I, Y\,II, La\,II, Ce\,II, Nd\,II and Eu\,II are selected from \cite{2018ApJ...865..129R}, and we revise the ($\rm{log\,gf}$) values by fitting the solar spectrum with LTE assumption. 
The detail informations are listed in Table \ref{tableA}. In addition, the hyperfine structure (HFS) plays an important role for strong absorption lines, and HFS of Sc \citep{2019ApJS..241...21L}, Cu \citep{2014ApJ...782...80S}, Mn \citep{2011ApJS..194...35D}, La \citep{2006ApJ...645..613I} and Eu \citep{2001ApJ...563.1075L} has been taken into account.



\subsection{Elemental abundances and their uncertainties}
For all elements, the abundances of our program stars were determined using the spectral synthesis method, and the one dimensional LTE MAFAGS opacity sampling (OS) models in plane-parallel atmospheres \citep{2004A&A...420..289G, 2009A&A...503..177G} have been adopted. It is noted by \citet{2017ApJ...845..151R} that the barium abundances are over-estimated by the standard LTE methods, 
our previous work \citep{2020ApJ...896...64L} has also presented that the non-local thermodynamic equilibrium (NLTE) effects can not be neglect, and the barium abundances can be over-estimated by up to 0.2 dex compared to the NLTE analysis. Thus, the NLTE effects have been considered in this work when deriving Barium abundances.

An interactive IDL code Spectrum Investigation Utility \citep[SIU,][]{Reetz91} is used to perform the line formation.   In order to fit the observed spectral lines, we handled the broadenings due to the rotation, macro-turbulence and instrument as one single Gauss profile to be convolved with the synthetic spectra. It need to be pointed out that our sample stars are not fast rotators.



\begin{figure*}[!t]
\centering
\includegraphics[width=0.97\textwidth]{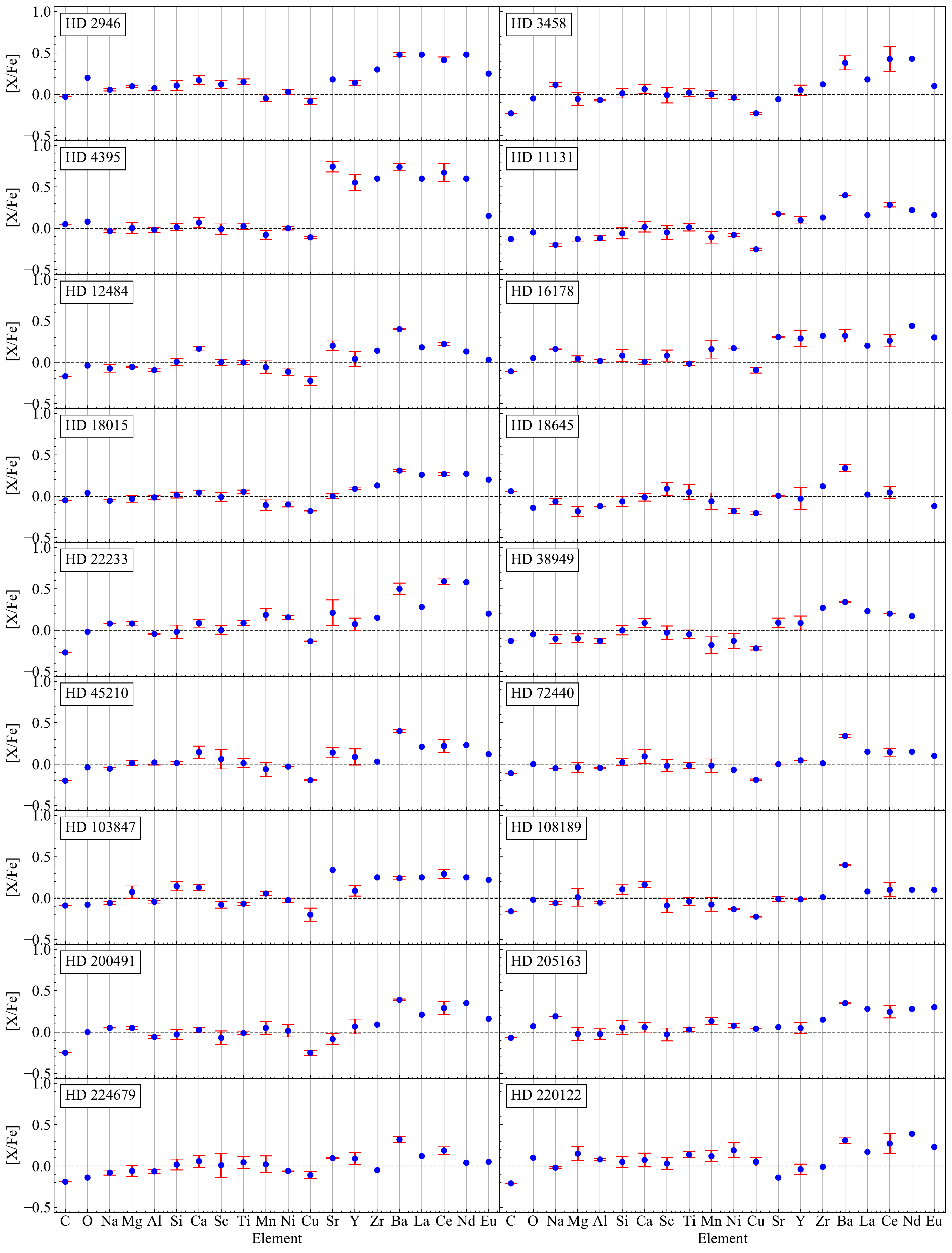}
\caption{\centering The elemental abundances of our 18 sample stars.}
\label{figure4}
\end{figure*}

The final abundances of 20 elements (C, O, Na, Mg, Al, Si, Ca, Sc, Ti, Mn, Ni, Cu, Sr, Y, Zr, Ba, La, Ce, Nd and Eu) are given in Table \ref{tableB} ([Sr/Fe] ratio represents the mean value of [Sr\,I/Fe] and [Sr\,II/Fe] ratios), and the uncertainties presented are the statistical standard deviation of different spectral lines for each element. In Figure \ref{figure4} we have plotted [X/Fe] v.s. the atomic number Z for the 18 stars in our sample.

It needs to be pointed out that only one line in our spectrum range is suitable to derive chemical abundance for each of the following six elements, e.g., C, Sr, Zr, La, Nd and Eu.  Due to the poor quality of the Sr\,II lines at 4077\,\AA\ and 4215\,\AA, we used the Sr\,I and Sr\,II lines at 4607 and 4161\,\AA, respectively, to derive the abundance of this element. 

The uncertainties in the  stellar parameters ($\mathrm{T_{eff}}$, log g, [Fe/H] and $\xi_{t}$) will lead to the errors of elemental abundances, and we select one typical star of HD\,16178 to investigate this impact. The errors in elemental abundances due to the uncertainties  of 100\,K in $\mathrm{T_{eff}}$, 0.2\,dex in log g, 0.1\,dex in [Fe/H] and 0.2\,km $s^{-1}$ in $\xi_{t}$ are presented in Table~\ref{table2}.
The last column lists the square roots of the quadratic sum of the errors associated with all the four factors. 
As shown in the Table, the uncertainties due to one of the stellar parameters are less than 0.10\,dex for most of the elements. The abundances of O and Sr are sensitive to temperature, while the strong lines of Ti, Ni, Sr and Ba are sensitive to micro-turbulence. 

\begin{table*}[htb]
\centering
\caption{ {\upshape Estimated uncertainties in abundance analysis for one example star of HD\,16178.}}
\label{table2}
 \setlength{\tabcolsep}{4mm} 
\begin{tabular}{lccccc}
\hline
\hline
$\Delta\,[X/Fe]$   &$\Delta\,T_{eff}$    &$\Delta\,log\,g$      &$\Delta\,[Fe/H]$       &$\Delta\,\xi_{t}$       &$\Delta\,Total$ \\
                   &+100 K                & +0.2 dex                & +0.1 dex                   & +0.2 $\rm{km}\,\rm{s^{-1}}$             &               \\
\hline                                                                                                                     
C\,I         &$-$0.07              &+0.07                  &$-$0.10                &+0.02              &+0.14           \\
O\,I         &$-$0.15              &+0.09                  &$-$0.09                &$-$0.02            &+0.20           \\
Na\,I        &+0.07                 &$-$0.02               &$-$0.10                &$-$0.04            &+0.13           \\
Mg\,I        &+0.06                 &$-$0.01               &$-$0.08                &$-$0.08            &+0.13           \\
Al\,I        &+0.08                 &+0.02                  &$-$0.07                &+0.01             &+0.10           \\
Si\,I        &$-$0.01              &+0.04                  &$-$0.07                &$-$0.03            &+0.09           \\
Ca\,I        &+0.05                 &$-$0.04               &$-$0.06                &$-$0.10            &+0.13           \\
Sc\,II        &+0.01                 &+0.10                  &$-$0.06                &$-$0.07           &+0.14           \\
Ti\,II        &+0.00                 &+0.07                  &$-$0.07                &$-$0.11           &+0.15           \\
Mn\,I        &+0.10                 &+0.01                  &$-$0.03                &$-$0.08           &+0.13           \\
Ni\,I        &+0.09                 &+0.03                  &$-$0.05                &$-$0.14           &+0.17           \\
Cu\,I        &+0.05                 &+0.01                  &$-$0.07                &$-$0.08           &+0.11           \\
Sr\,I        &+0.18                 &+0.00                  &$-$0.08                &$-$0.12           &+0.23           \\
Sr\,II        &+0.06                 &+0.10                  &$-$0.04                &$-$0.07           &+0.14           \\
Zr\,II        &$-$0.04              &+0.09                  &$-$0.07                &$-$0.06            &+0.13           \\
Ba\,II        &+0.03                 &$-$0.13               &$-$0.05                &$-$0.12            &+0.18           \\
Eu\,II        &$-$0.02              &+0.08                  &$-$0.05                &$-$0.03            &+0.10           \\
Y\,II         &+0.00                 &+0.08                  &$-$0.04                &$-$0.07           &+0.11           \\
La\,II        &+0.04                 &+0.09                  &$-$0.04                &+0.01             &+0.11           \\
Ce\,II        &+0.03                 &+0.08                  &$-$0.03                &$-$0.08           &+0.12           \\
Nd\,II        &+0.05                 &+0.07                  &$-$0.04                &$-$0.06           &+0.11           \\
\hline                                                                                                     
\hline        

\end{tabular}
\end{table*}

\subsection{Comparing to previous works}
\begin{figure}
\centering
\includegraphics[width=0.7\linewidth]{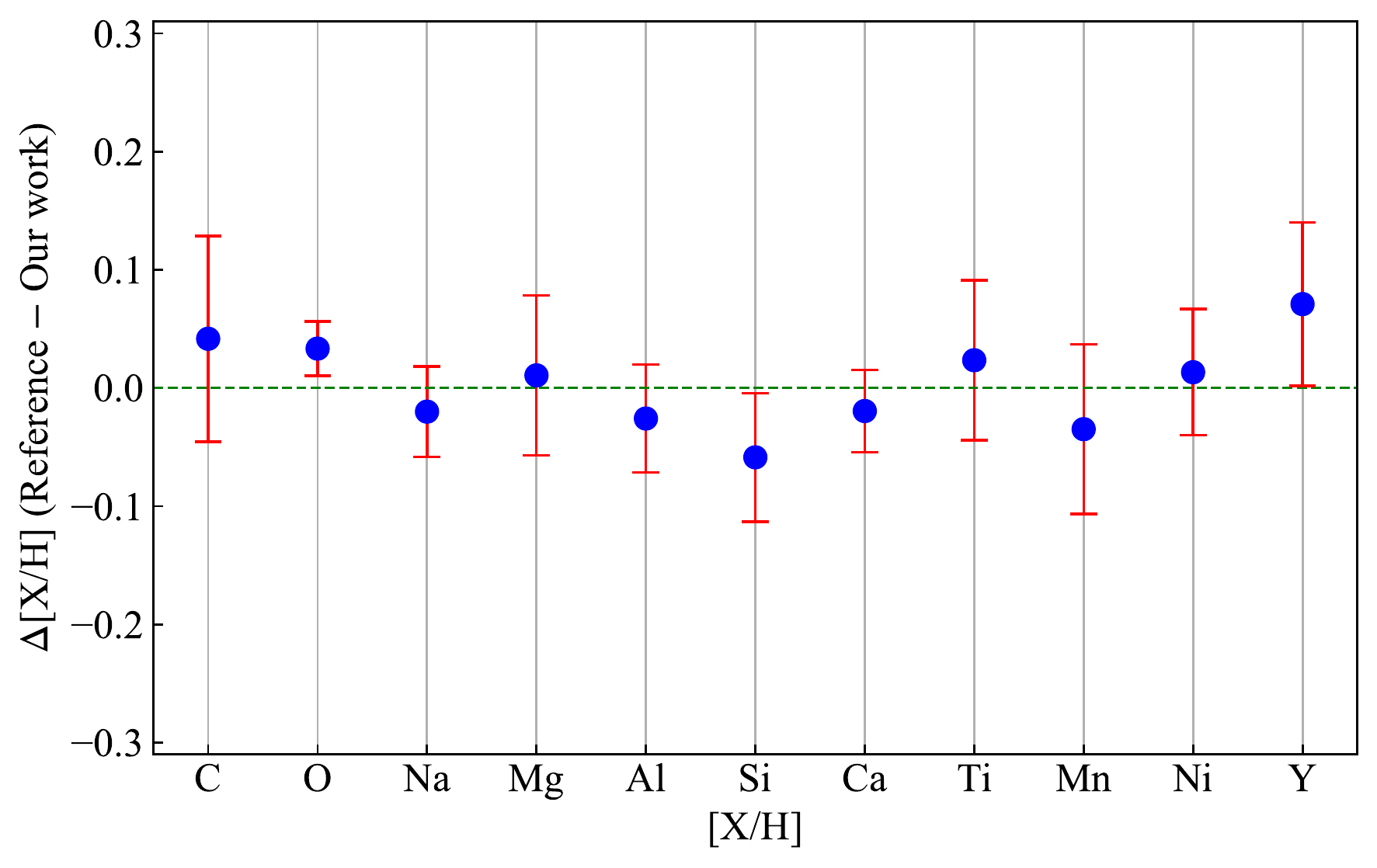}
\caption{The differences of elemental abundances between ours and \cite{2016ApJS..225...32B}. Only the objects with $\Delta$$\rm{T_{eff}}$ $<$ 100\,K, $\Delta$$\rm{log\ g}$ $<$ 0.1\,dex and $\Delta$[Fe/H]) $<$ 0.1\,dex have been selected, which contains 10 common stars.}
\label{figure5}
\end{figure}

In order to validate our results on the element abundances, we compared them with those from previous works. \cite{2016ApJS..225...32B} gave the abundances of 15 elements, which have 13 common elements with our study. To avoid the uncertainties due to different atmospheric stellar parameters adopted, we compared the abundance results only for the ten objects 
with $\Delta\rm{T_{eff}}<$ 100\,K, $\Delta\rm{log\ g}<$ 0.1\,dex and $\Delta$[Fe/H] $<$ 0.1\,dex. 
The comparison results are presented in Figure \ref{figure5}, and it shows that the mean deviation is lower than 0.1\,dex for all elements.

\begin{figure}
\centering
\includegraphics[width=0.7\linewidth]{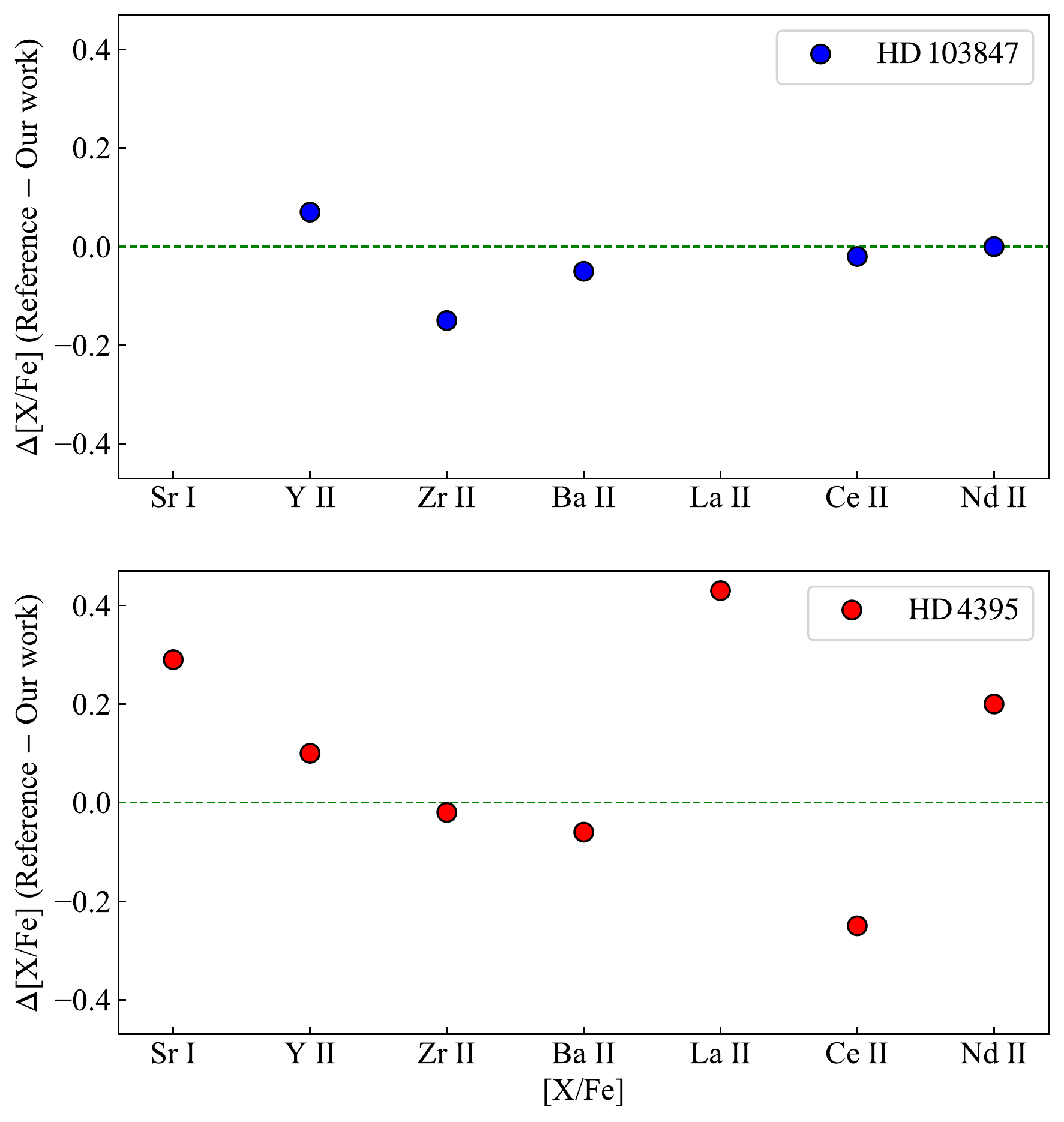}
\caption{The comparison of the difference in the s-process elemental abundances for two common stars of HD\,103847 and HD\,4395 studied by \cite{2012A&A...547A..13T} and \cite{2014MNRAS.440.1095K}, respectively.}
\label{figure6}
\end{figure}

We notice that the abundances of  s-process elements for two stars, HD\,103847 and HD\,4395, have been determined by \cite{2012A&A...547A..13T} and \cite{2014MNRAS.440.1095K}, respectively. The differences between our values and those of the references are presented in Figure \ref{figure6} (also see Table\,\ref{table3}). As shown in the figure, our values have a good agreement with those of \cite{2012A&A...547A..13T} for HD\,103847, while they are large discrepancy of Sr, La, Sr and Nd for HD\,4395 \citep{2014MNRAS.440.1095K}. 
We find that the stellar atmosphere parameters adopted for HD\,4395 are very similar for ours and these from \citet{2014MNRAS.440.1095K}. It is noted that the Sr abundance are derived from the only Sr\,I line at 4607.33\,$\rm{\AA}$ for both works, and the corresponding difference in $\rm{log\,gf}$ values is $-$0.62 ($\rm{{log\,gf}_{Ref.}=-0.57}$, $\rm{{log\,gf}_{our}=0.28}$).  Thus, the deviation can partially be explained by the different $\rm{log\,gf}$ values adopted. Moreover, the Sr abundance of 13 sample stars are also derived with the Sr\,II line at 4161\,\AA, and the mean deviation from the [Sr\,I/Fe] and [Sr\,II/Fe] 
is 0.06\,dex.  For the abundance of Ce, we have one common Ce\,II line at 5330.56\,\AA, and the difference in $\rm{log\,gf}$ value is $-$0.46 ($\rm{{log\,gf}_{Ref.}=-0.76}$, $\rm{{log\,gf}_{our}=-0.30}$).
We also check the equivalent widths of this line, and the values are 12.7\,m\AA\ (Ref.) and 15.6\,m\AA\ (this work), respectively. Our Ce abundance is obtained from four Ce\,II lines, and they give consistent abundance results, while theirs are based on only two lines. The abundance differences in La and Nd are hard to discuss, as there are no common lines for these two works.

\begin{table}[htb]
\begin{threeparttable}
\centering
\caption{ {\upshape The comparison of the abundances of s-process elements for HD\,103847 and HD\,4395.}}
\label{table3}
 \setlength{\tabcolsep}{3mm} 
\begin{tabular}{lrrrrrrrc}
\hline
\hline
Star Name& [Sr\,I/Fe] & [Y\,II/Fe] & [Zr\,II/Fe] & [Ba\,II/Fe] & [La\,II/Fe] & [Ce\,II/Fe] & [Nd\,II/Fe] & Remarks\\
\hline
HD\,103847&0.34	&	0.09	&	0.25&		0.24&		0.25&		0.29&		0.25& Our work
\\
&$-$  &0.16  &0.10   &0.19  &$-$   &0.27 &0.25 & T12
\\
HD\,4395&  0.79	&	0.55	&	0.60  &0.86&		0.60&		0.67&		0.60& Our work\\
        & 1.08  &0.65       &0.58     &0.79        &1.03     &0.42         &0.80& K14\\
\hline 
\end{tabular}
\begin{tablenotes}
\item
\centering
Notes. T12: \cite{2012A&A...547A..13T}, K14: \cite{2014MNRAS.440.1095K}.
\end{tablenotes}
\end{threeparttable}
\end{table}

\section{Discussion}
\subsection{Identification of Ba stars}

As previously mentioned, \citet{2016MNRAS.459.4299D} suggested a minimal value of [s/Fe]\,=\,0.25 as a criterion to identify Ba stars. According to this rule, four objects in our sample can be classified as Ba stars including one known CH subgiant of HD\,4395 ([s/Fe]\,=\,0.65), and three are newly discovered e.g. HD\,2946 ([s/Fe]\,=\,0.35), HD\,16178 ([s/Fe]\,=\,0.31) and HD\,22233 ([s/Fe]\,=\,0.36). Among these three new Ba stars, HD\,16178 and HD\,22233 are Ba giants with $log\,g$\,=\,3.11 and 2.72, respectively, while HD\,2946 is a Ba subgiant with $log\,g$\,=\,3.70. For the rest of fourteen stars, they all show  Ba enhancement of [Ba/Fe] $>$ 0.3 except for HD\,103847, the later has a [s/Fe] value of 0.24\,dex, which is close to the standard of Ba stars. Thus, these fourteen stars, which contain six dwarfs and eight giants, are added as a reference in the following discussion.



\subsection{Carbon abundances}

The first Ba stars identified by \cite{1951ApJ...114..473B} is enhancement of carbon, and according to the classical Ba star scenario, carbon can be produced in the form of $\rm{^{12}C}$ by shell He burning in TP-AGB stage {\citep{1997ApJ...476L..89P}. Figure~\ref{figure7} shows the ratio of [C/Fe] vs. metallicity,  and there are no significant enrichment of carbon found in our sample stars, including the known CH subgiant of HD\,4395 ([C/Fe] = 0.05). The  distribution of  [C/Fe] of the program stars follows the galactic field dwarfs. 

\subsection{The abundances of $\alpha$, odd and iron group elements}

As shown in Figure~\ref{figure4}, most sample stars have similar chemical abundances in Na, Al, Sc $\alpha$- and iron-peak elements (Na, Mg, Al, Si, Ca, Sc, Ti, Mn and Ni). 
We notice that the abundances of copper are slightly lower, which may be explained by the NLTE effect. According to former research \citep{2014ApJ...782...80S, 2019ApJ...875..142X}, the NLTE effect for copper is about $+$0.05\,dex.


\begin{figure}
\centering
\includegraphics[width=0.7\linewidth]{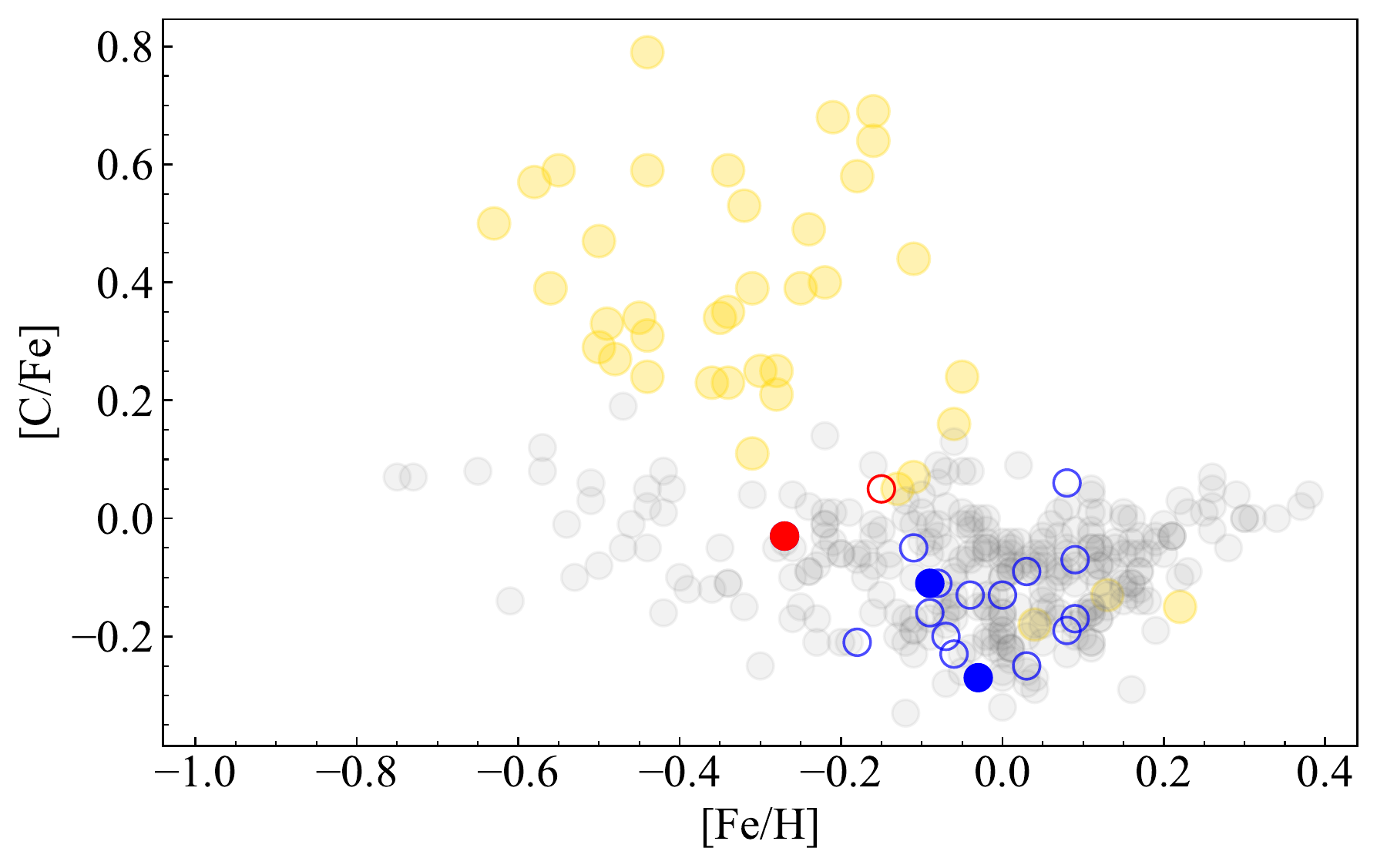}
\caption{The [C/Fe] ratio plotted against metallicity. The filled blue and red circles are our newly discovered Ba giants and subgiant, respectively. The open blue and red circles are the known CH subgiant HD\,4395 and 14 comparison stars, respectively.
The grey dots are the field FGK dwarfs from \cite{2015A&A...580A..24D}, and the yellow dots are for the known Ba dwarfs and CH subgiants from \cite{2018MNRAS.474.2129K}.}
\label{figure7}
\end{figure}

\begin{figure}
\centering
\includegraphics[width=0.7\linewidth]{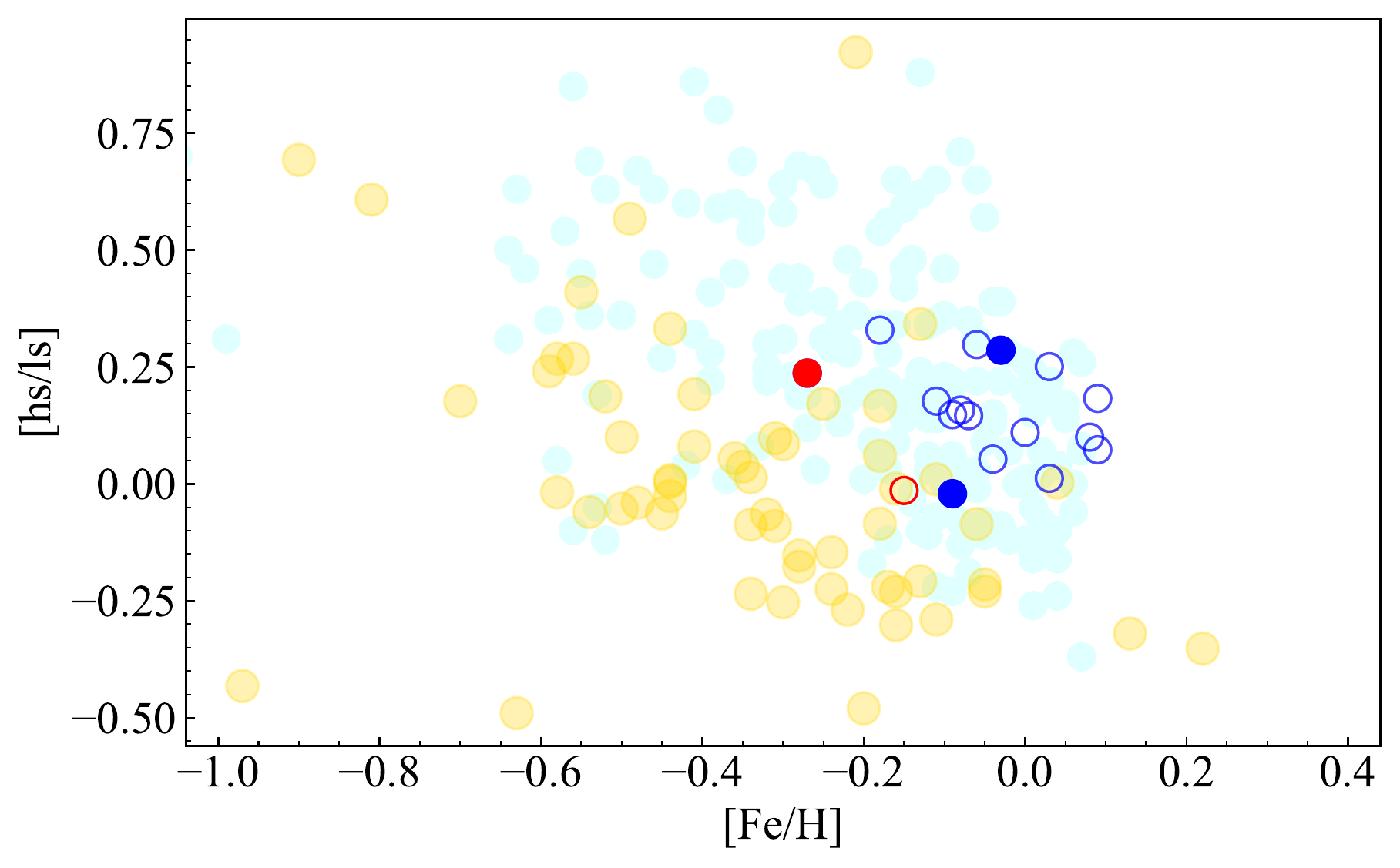}
\caption{The [hs/ls] ratios plotted against metallicity. The blue and red circles have the same meaning as Figure~\ref{figure7}. The yellow dots are the known Ba dwarfs and CH subgiants from \cite{2018MNRAS.474.2129K}, and the cyan dots is the known Ba giants from \cite{2016MNRAS.459.4299D}. }
\label{figure8}
\end{figure}

\subsection{The abundances of s-process elements}


The [hs/ls] ratio is a powerful tool to diagnose the efficiency of s-process \citep{1991ApJS...77..515L}, here, [hs] and [ls] are the mean abundances of the heavy (Ba, La, Ce and Nd) and light s-process elements (Sr, Y and Zr), respectively. Recently, \cite{2018MNRAS.474.2129K} found that the [hs/ls] ratios show an anti-correlation with metallicity for CH subgiant and dwarf Ba stars. Figure~\ref{figure8} presents the ratios of [hs/ls] as a function of metallicity, and for comparison we also include the samples from \cite{2016MNRAS.459.4299D} and \cite{2018MNRAS.474.2129K}. 
As shown in the bottom panel of Figure~\ref{figure8}, the [hs/ls] ratios of the three new Ba stars  are consistent with those of the known Ba stars, and the subgiant Ba star of HD\,2946 tends having higher [hs/ls] value than that of the most known Ba dwarfs at the same metallicity. As highlighted in \cite{2018MNRAS.474.2129K}, an anti-correlation existed between [hs/ls] and metallicity in the sample of Ba dwarfs and CH subgiants, while this relevance is not obvious for Ba giants.   
In addit{}ion, with regard to the fourteen Ba candidates, no matter giants or dwarfs, the distribution of [hs/ls] ratios is similar to the known Ba giants.
{}
\subsection{The abundances of r-process element Eu}

It has been found that europium mainly comes from the r-process \citep[e.g.][]{2006A&A...448..557C}, and the [Eu/Fe] ratio is a diagnostic to determine whether the r-process dominated the nucleosynthesis at a given moment of the evolution of the Galaxy. At very low metallicities, the r-process is expected to be the dominate heavy element production, since the massive stars first explode in the form of nuclear collapse and enrich the interstellar medium (ISM) before the AGB stars participate in the main s-process. In this work, we also derived the Eu abundances, and the results are presented in Figure~\ref{figure9}. We can see that the Eu abundances of our Ba stars are similar to these of the thick-disk stars.

\begin{figure}
\centering
\includegraphics[width=0.7\linewidth]{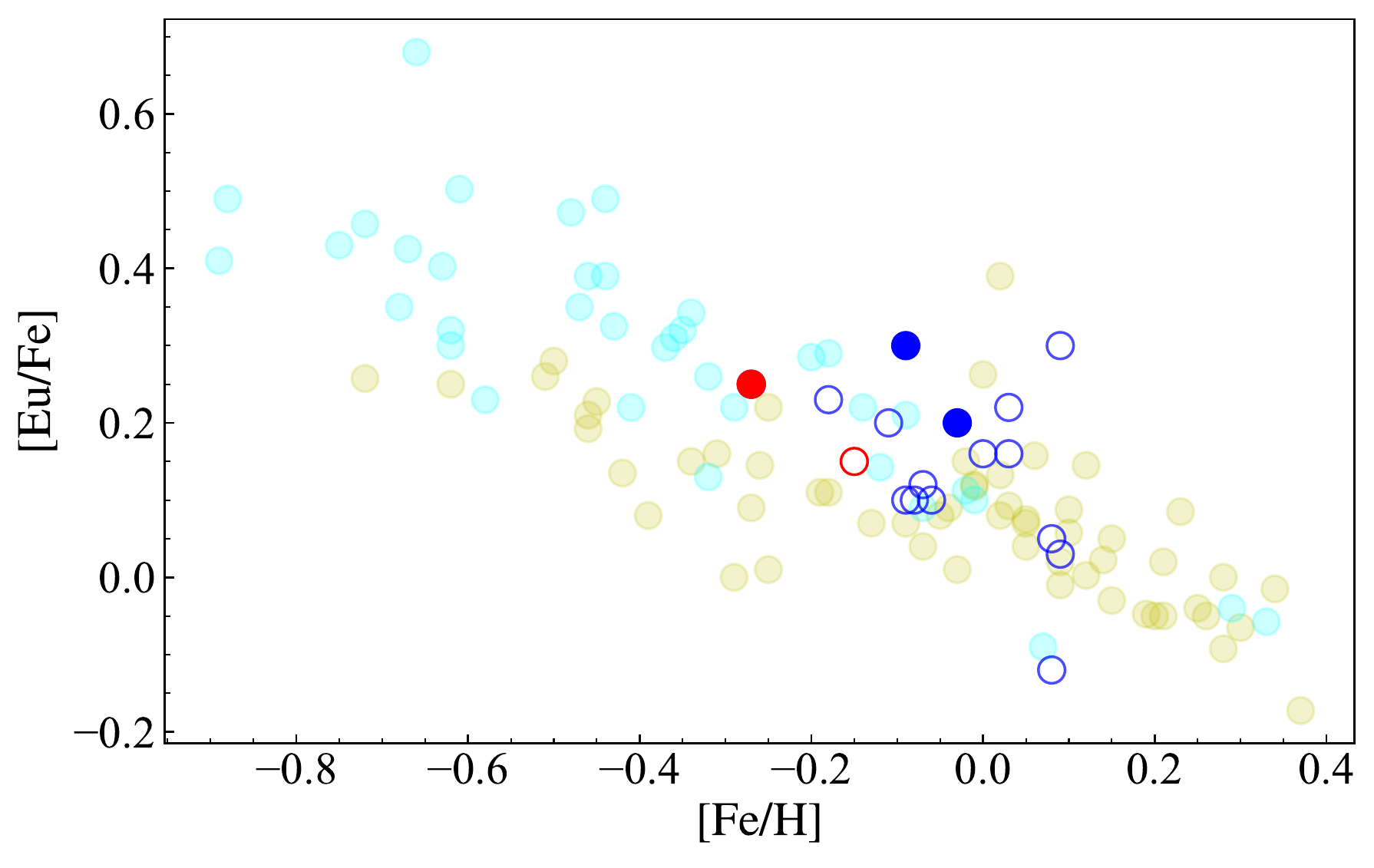}
\caption{The [Eu/Fe] ratio against metallicity. The blue and red circles are the same as Figure~\ref{figure7}, while the yellow and cyan dots are the thin- and thick-disk stars from \cite{2005A&A...433..185B}}
\label{figure9}
\end{figure}

\subsection{Radial velocities of the four Ba stars}
In this work, we acquired the high precise radial velocities of our four Ba stars by using the public online code of HIRES Precision Radial Velocity Pipeline \footnote{\url{https://caltech-ipac.github.io/hiresprv/index.html}} \citep{1992PASP..104..270M,1996PASP..108..500B,2010ApJ...721.1467H}. The typical uncertainty of radial velocities is around 1-3\,m\,s$^{-1}$. 

\begin{figure*}[!t]
\centering
\includegraphics[width=1.0\textwidth]{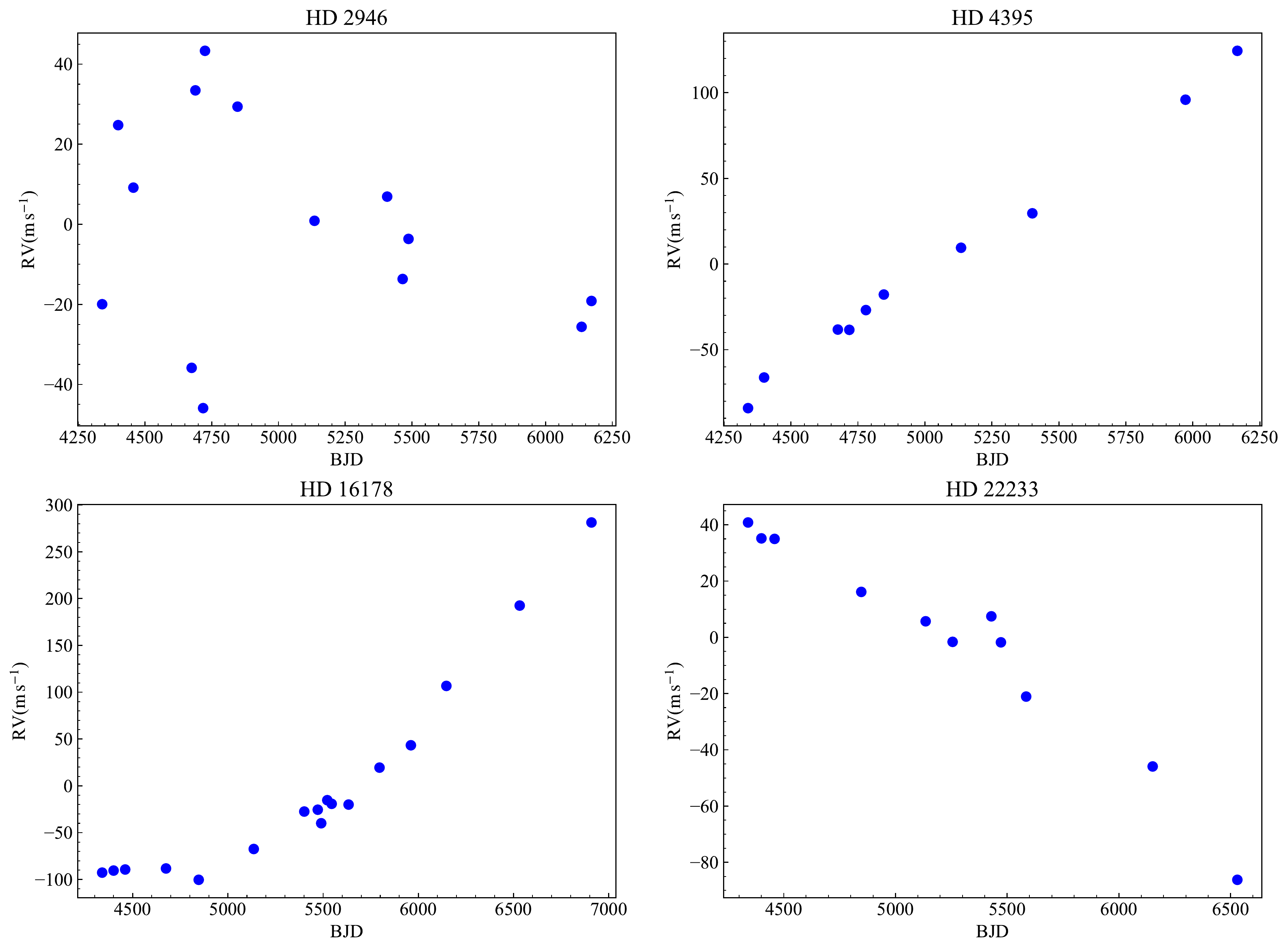}
\caption{\centering The radial velocities of our four Ba stars.}
\label{figure10}
\end{figure*}

The variations of radial velocities for the four Ba stars are presented in Figure~\ref{figure10}. It can be found that three stars (HD\,4395, HD\,16178 and HD\,22233) have obvious secular linear trends on radial velocities, suggesting the existences of long-period companions around these stars. The acceleration ($|\dot{\gamma}|$) are between 20 to $70\,\mathrm{m\,s^{-1}\,yr^{-1}}$. It is difficult to estimate the minimum mass of the companions with the existing radial velocity data. Their companions can be either planet, brown dwarf, or low-mass stars. Long-term monitoring are essential to uncover their properties. For HD\,2946, no linear radial velocity trends are observed. We also calculate the Generalized Lomb-Scargle periodogram \citep{2009A&A...496..577Z} of the radial velocity data, and did not find any significant peak with False Alarm Probability (FAP) less than 10$^{-3}$, which means there is no clue of hosting stellar or substellar companions. However, binarity of Ba subgiant HD\,2946 cannot be excluded. High spatial-resolution imaging will be helpful to reveal their possible companions.

\section{Conclusions}
In this work, based on the high resolution and high S/N spectra from Keck/HIRES, we have determined the abundances of 20 elements (C, O, Na, Mg, Al, Si, Ca, Sc, Ti, Mn, Ni, Cu, Sr, Y, Zr, Ba, La, Ce, Nd and Eu) for 18 Ba candidates. The parameter space for our sample stars are in the range of 4880 $\leq$ $\rm{T_{eff}}$ $\leq$ 6050\,K, 2.56 $\leq$ log $g$ $\leq$ 4.53\,dex and $-$0.27 $\leq$ [Fe/H] $\leq$ 0.09\,dex. 
The conclusions are as follows.

\begin{itemize}
\item{We confirm that four of our 18 Ba candidates are Ba stars, and three of them are newly discovered including one Ba subgiant of HD\,2946 and two Ba giants of HD\,16178 and HD\,22233.}
\item {Three of our four Ba stars, HD\,4395, HD\,16178 and HD\,22233,  have clear evidences of hosting stellar or sub-stellar companions in long-term radial velocity data, and the binarity for the Ba subgiant HD\,2946 cannot be excluded.} 
\item{The distribution of the [hs/ls] ratios for all of four Ba stars is similar to that of the known Ba objects, and our results support the suggestion that there is an anti-correlation between the [hs/ls] ratio and metallicity for CH subgiant and dwarf Ba stars}.
\item{ No significant enrichment of carbon is found for our Ba stars, including the known CH subgiant HD\,4395.}
\end{itemize}
Combining more observations on the radial velocities and photometric in longer time range, we could further solve the orbit parameters of the binaries. In addition, it is important to search for more metal-poor Ba stars ([Fe/H] $<$ $-$0.6) and metal-rich Ba dwarfs ([Fe/H $>$ $-$0.1]) to investigate the relation between the [hs/ls] ratios and metallicities.

\normalem
\begin{acknowledgements}
This research is supported National Key R\&D Program of China No.2019YFA0405502, the National Natural Science Foundation of China under grant Nos. 12090040, 12090044, 11833006, 12022304, 11973052, 11973042, U2031144 and U1931102. This work is also supported by the Astronomical Big Data Joint Research Center, co-founded by the National Astronomical Observatories, Chinese Academy of Sciences and Alibaba Cloud. H.-L.Y. acknowledges support from the Youth Innovation Promotion Association of the CAS (id. 2019060). This work is also partially supported by the Open Project Program of the Key Laboratory of Optical Astronomy, National Astronomical Observatories, Chinese Academy of Sciences. 
The data presented herein were obtained at the W. M. Keck Observatory, which is operated as a scientific partnership among the California Institute of Technology, the University of California and the National Aeronautics and Space Administration. The Observatory was made possible by the generous financial support of the W. M. Keck Foundation. This work has made use of data from the European Space Agency (ESA) mission {\it Gaia} (\url{https://www.cosmos.esa.int/gaia}), processed by the {\it Gaia} Data Processing and Analysis Consortium (DPAC, \url{https://www.cosmos.esa.int/web/gaia/dpac/consortium}). Funding for the DPAC has been provided by national institutions, in particular the institutions participating in the {\it Gaia} Multilateral Agreement. his research has also made use of the Washington Double Star Catalog maintained at the U.S. Naval Observatory.
 
\end{acknowledgements}

 


\bibliographystyle{raa}
\bibliography{s-process}

\begin{appendix}

\renewcommand{\thetable}{A}
\setcounter{table}{0}
\begin{table}[b]

\centering
\caption{ {\upshape Absorption lines used for abundance determination.}}
\label{tableA}
 \setlength{\tabcolsep}{10mm}                                         
\begin{tabular}{lllr}                                                
\hline                                                                
\hline                                                                 
Ion & $\rm{\lambda (\AA)}$ & $\chi (eV)$ & $\rm{log\ gf}$ \\          
\hline              	                                                  
   C\,I	   			    & 5380.33  &  7.68       &  $-$1.62  \\           
  O\,I	    		   & 6300.30  &  0       &   $-$9.78       \\                         
  Na\,I	  	   	     & 6154.23  &  2.10       &   $-$1.55  \\          
  						       & 6160.75  &  2.10       &   $-$1.25  \\          
  Mg\,I    			   & 4571.09   &  0.00       &   $-$5.47         \\  
  						      & 4730.03   &  4.34       &   $-$2.20          \\  
  						      & 5711.07   &  4.34       &   $-$1.64          \\  
  Al\,I    		    	& 6696.03   &  3.14       &   $-$1.51        \\    
  						      & 6698.67   &  3.14       &   $-$1.87        \\    
  Si\,I   		   		& 5690.42    &  4.71      &   $-$1.74        \\    
  						      & 5701.10    &  4.71      &   $-$1.96        \\    
  						      & 5772.15    &  5.08      &   $-$1.62        \\    
  						      & 6142.48    &  5.62      &   $-$1.48      \\      
  						      & 6145.02    &  5.62      &   $-$1.39      \\      
  						      & 6155.13    &  5.62      &   $-$0.78      \\      
  						      & 6237.32    &  5.61      &   $-$1.08      \\      
  						      & 6243.81    &  5.62      &   $-$1.29      \\      
  						      & 6244.47    &  5.62      &   $-$1.29        \\    
  Ca\,I    				  & 5512.98    &  2.93      &   $-$0.46      \\      
   						      & 5867.57    &  2.93      &   $-$1.57       \\     
   						      & 6161.29    &  2.51      &   $-$1.27       \\     
   						      & 6166.44    &  2.51      &   $-$1.14       \\     
   						      & 6169.06    &  2.51      &   $-$0.80       \\     
   						      & 6169.56    &  2.53      &   $-$0.48       \\     
  Sc\,II	    			& 5031.02    &  1.35      &   $-$0.41      \\ 
  						      & 5641.00    &  1.49      &   $-$0.99      \\
  						      & 5657.90    &  1.50      &   $-$0.54      \\   
  						      & 5669.04    &  1.49      &   $-$1.10      \\      
  						      & 5684.19    &  1.50      &   $-$1.03      \\  
  Ti\,II	  				& 4470.85     &1.16       &   $-$2.02      \\       
  						     & 4583.41     &1.16        &   $-$2.84      \\       
  						     & 4708.66     &1.24        &   $-$2.35      \\      
  						     & 5336.79     &1.58        &   $-$1.60      \\      
  						     & 5418.77     &1.58        &   $-$2.13      \\      
  Mn\,I     			 & 4761.51     &2.95        &   $-$0.27      \\      
   						     & 4765.85     &2.94        &   $-$0.09      \\	     
   						     & 6013.51     &3.07        &   $-$0.35      \\      
   						     & 6021.82     &3.07        &   $-$0.05  \\ 
           
\hline
\end{tabular}
\end{table}

\begin{table}[b]

\centering
 \setlength{\tabcolsep}{12mm}                                       
\begin{tabular}{lllr}                                              
\hline                                                              
\hline                                                              
Ion & $\rm{\lambda (\AA)}$ & $\chi (eV)$ & $\rm{log\ gf}$ \\        
\hline                                                              
 Fe\,I      & 4574.22     &2.28 	  &$-$2.37              	 \\	
								 & 5242.50     &3.63    &$-$0.83                 \\	
								 & 5295.32     &4.42    &$-$1.54               	\\	
								 & 5379.58     &4.15    &$-$1.35               \\	  
								 & 5576.10     &3.43    &$-$0.76                  \\
								 & 5633.95     &4.99    &$-$0.19                 \\ 
								 & 5638.27     &4.22    &$-$0.70               \\   
								 & 5662.52     &4.18    &$-$0.45        	   \\     
								 & 5679.02     &4.65    &$-$0.68                 \\	
								 & 5705.48     &4.30    &$-$1.44                 \\	
								&  5814.81   &  4.28      & $-$1.85         \\      
								&  5852.22   &  4.55      & $-$1.20        \\       
								&  5855.08   &  4.61      & $-$1.55        \\       
								&	 5916.25   &  2.45      & $-$2.87       \\        
								&	 5930.18   &  4.65      & $-$0.18       \\        
								&  6024.06   &  4.55      &   $+$0.08      \\       
								&  6056.01   &  4.73      &   $-$0.36        \\     
								&  6065.49   &  2.61      &   $-$1.41        \\     
								&  6079.01   &  4.65      &   $-$0.97        \\     
								&  6127.90   &  4.14      &   $-$1.36      \\       
								&  6151.62   &  2.18      &   $-$3.26      \\       
								&  6165.36   &  4.14      &   $-$1.44      \\       
								&  6173.34   &  2.22      &   $-$2.82      \\       
								&  6180.20   &  2.73      &   $-$2.66      \\       
								&  6200.32   &  2.61      &   $-$2.32      \\       
								&  6213.43   &  2.22      &   $-$2.45      \\       
								&  6232.64   &  3.65      &   $-$1.12      \\       
								&  6252.55   &  2.40      &   $-$1.58      \\       
                &  6393.61   &  2.43      &   $-$1.43      \\       
                &  6593.87   &  2.44      &   $-$2.28      \\       
                &  6609.11   &  3.56      &   $-$2.58      \\       
                &  6733.15	 &  4.64      &   $-$1.44      \\       
                &  6750.15   &  2.42      &   $-$2.54      \\       
   Fe\,II      &  4508.28   &  2.86      &   $-$2.37      \\       
								&  4520.22   &  2.81      &   $-$2.60      \\       
								&  5197.58   &  3.22      &   $-$2.54      \\       
								&  5234.63   &  3.22      &   $-$2.24      \\       
								&  5264.81   &  3.22      &   $-$3.06      \\       
								&  5414.07   &  3.22      &   $-$3.57      \\       
								&  5425.26   &  3.20      &   $-$3.28      \\       
								&  5991.38   &  3.15      &   $-$3.60      \\       
								&  6084.11   &  3.20      &   $-$3.83      \\       
								&  6149.25   &  3.89      &   $-$2.74      \\       
								&  6247.56   &  3.89      &   $-$2.56      \\       
								&  7711.73   &  3.90      &   $-$2.58      \\       
\hline
\end{tabular}
\end{table}

\begin{table}[b]
\centering
  \setlength{\tabcolsep}{12mm} 
\begin{tabular}{lllr}
\hline
\hline
Ion & $\rm{\lambda (\AA)}$ & $\chi (eV)$ & $\rm{log\ gf}$  \\      
\hline  
 Ni\,I     &  6300.34   &  4.27      &  $-$2.11      \\                                                                
		   &  6643.63   &  1.68      &  $-$2.22      \\    
 			     &  6767.77   &  1.82      &  $-$2.14      \\    
 Cu\,I     &  5105.54   &  1.39      &  $-$1.52      \\    
 			   	 &  5218.20   &  3.82      &  0.48      \\       
 Sr\,I     &  4607.33   &  0.00      &  0.28        \\   
 Sr\,II    &  4161.79   &  2.94      & $-$0.50        \\  
 Y\,II     &  5087.42   &  1.08      &  $-$0.16       \\	 
 		       &  5402.77   &  1.84      &  $-$0.31         \\ 
 		       &  5544.61   &  1.74      &  $-$0.83        \\	 
 Zr\,II    &  5112.27   &  1.66      &  $-$0.85        \\	 
 Ba\,II    &  5853.67   &  0.60      &  $-$1.00        \\	 
 		     	 &  6141.71   &  0.70      &  $-$0.08       \\	     
 La\,II    &  6390.48   &  0.32      &  $-$1.41        \\	 
 Ce\,II    &  4562.36   &  0.48      &  0.21       \\      
 		       &  4572.28   &  0.68      &  0.22        \\     
 		       &  4628.16   &  0.52      &  0.14         \\    
 		       &  5330.56   &  0.87      &  $-$0.40         \\ 
 Nd\,II    &  5319.81   &  0.55      &  $-$0.14       \\   
 Eu\,II    &  6645.06   &  1.38      &  0.12     \\	   
\hline
\end{tabular}
\end{table}

\begin{landscape}
\vspace*{\fill}
\renewcommand{\thetable}{B}
\setcounter{table}{0}
\begin{table}
\centering
\caption{The elemental abundances of our sample stars.}
\setlength{\tabcolsep}{0.5mm} 
\begin{tabular}{lrcrcrcrcrcrcrcrcrcrc}
\hline
\hline
Star Name       &[C/Fe]	        &error	  &[O/Fe]	       &error	       &[Na/Fe]	  &error	     &[Mg/Fe]	         &error	   &[Al/Fe]	    &error	  &[Si/Fe]	 &error	  &[Ca/Fe]	  &error	 &[Sc/Fe]	   &error	  &[Ti/Fe]	     &error	  &[Mn/Fe]	   &error   \\ 
\hline                                                                                                                                                                                                                                                           
HD\,2946        & $-$0.03      &$-$	      &0.13        &0.059        &0.06         &0.015	      &0.10          &0.012	   &0.08        &0.025	  &0.11      &0.058	     &0.17     &0.056	    &   0.12   &0.047	   &0.15        &0.036	  &$-$0.05     &0.040  \\  
HD\,3458        & $-$0.23      &$-$	      &0.18        &0.071        &0.12         &0.025	      &$-$0.06	     &0.079	   &$-$0.07     &0.010 	  &0.03      &0.057	     &0.06     &0.054	    &$-$0.01   &0.094	   &0.02        &0.051	  &0.00        &0.050  \\  
HD\,4395        & 0.05         &$-$	      &0.10        &0.043        &$-$0.04      &0.015	      &0.00          &0.066	   &$-$0.02     &0.030 	  &0.01      &0.040 	   &0.07     &0.063	    &$-$0.01   &0.063	   &0.02        &0.037	  &$-$0.08     &0.055  \\  
HD\,11131       & $-$0.13      &$-$	      &0.09        &0.111        &$-$0.20      &0.020 	    &$-$0.13 	     &0.024	   &$-$0.12     &0.030 	  &$-$0.06   &0.067	     &0.02     &0.062	    &$-$0.05   &0.083 	 &0.01        &0.044	  &$-$0.11     &0.071  \\  
HD\,12484       & $-$0.17      &$-$	      &0.04        &0.090        &$-$0.08      &0.045	      &$-$0.06	     &0.005	   &$-$0.10     &0.015	  &0.00      &0.042	     &0.16     &0.026	    &   0.00   &0.035	   &0.00        &0.025	  &$-$0.06     &0.076  \\  
HD\,16178       & $-$0.11      &$-$	      &0.23        &0.111        &0.16         &0.010 	    &0.04          &0.034	   &0.02        &0.015	  &0.08      &0.075	     &0.01     &0.031	    &   0.08   &0.067	   &$-$0.02     &0.025	  &0.16        &0.108  \\  
HD\,18015       & $-$0.05      &$-$	      &0.16        &0.083        &$-$0.06      &0.015	      &$-$0.03       &0.039	   &$-$0.02     &0.025	  &0.01      &0.035	     &0.04 	   &0.029	    &$-$0.01   &0.052	   &0.05        &0.021	  &$-$0.11     &0.063  \\  
HD\,18645       & 0.06         &$-$	      &0.13        &0.102        &$-$0.07      &0.035	      &$-$0.18	     &0.059	   &$-$0.12     &0.000    &$-$0.06   &0.057	     &-0.01	   &0.045	    &   0.09   &0.080	   &0.05        &0.091    &$-$0.06     &0.101  \\  
HD\,22233       & $-$0.27      &$-$	      &$-$0.04     &0.076        &0.08         &0.000    	  &0.08          &0.028	   &$-$0.05     &0.005	  &$-$0.02   &0.082	     &0.09     &0.048	    &   0.00   &0.053	   &0.09        &0.032	  &0.19        &0.074  \\  
HD\,38949       & $-$0.13      &$-$	      &0.05        &0.069        &$-$0.11      &0.055	      &$-$0.10  	   &0.054	   &$-$0.13 	  &0.030 	  &0.00     &0.055	     &0.09     &0.055	    &$-$0.03   &0.081	   &$-$0.05     &0.053	  &$-$0.18     &0.100  \\  
HD\,45210       & $-$0.20      &$-$	      &0.01        &0.024        &$-$0.06      &0.015	      &0.01          &0.029	   &0.02        &0.030 	  &0.01      &0.018	     &0.15     &0.073	    &   0.06   &0.119	   &0.01        &0.055	  &$-$0.06     &0.084  \\  
HD\,72440       & $-$0.11      &$-$	      &0.01        &0.038        &$-$0.05      &0.000    	  &$-$0.04       &0.062	   &$-$0.05     &0.005    &0.02      &0.042	     &0.10     &0.087	    &$-$0.02   &0.071    &$-$0.02     &0.039	  &$-$0.02     &0.081  \\  
HD\,103847      & $-$0.09      &$-$	      &0.05        &0.029        &$-$0.06      &0.020 	    &0.07 	       &0.073	   &$-$0.05     &0.015	  &0.14      &0.056	     &0.13     &0.035	    &$-$0.08   &0.041	   &$-$0.07     &0.019	  &0.06        &0.027  \\  
HD\,108189      & $-$0.16      &$-$	      &0.23        &0.093        &$-$0.06      &0.020 	    &0.01  	       &0.108	   &$-$0.06     &0.015	  &0.11      &0.063	     &0.16     &0.037	    &$-$0.09	 &0.086	   &$-$0.04     &0.046	  &$-$0.08     &0.087  \\  
HD\,200491      & $-$0.25      &$-$	      &$-$0.01     &0.071        &0.05         &0.000    	  &0.05          &0.016	   &$-$0.06     &0.020 	  &$-$0.03   &0.063	     &0.03     &0.035	    &$-$0.07   &0.083	   &$-$0.01     &0.015	  &0.05        &0.079  \\  
HD\,205163      & $-$0.07      &$-$	      &0.29        &0.127        &0.19         &0.000    	  &$-$0.02       &0.079	   &$-$0.03     &0.065	  &0.05      &0.085	     &0.06     &0.058	    &$-$0.03   &0.078	   &0.03        &0.021	  &0.13        &0.045  \\  
HD\,220122      & $-$0.21      &$-$	      &0.15        &0.057        &$-$0.02      &0.010 	    &0.15          &0.086	   &0.08        &0.010    &0.05      &0.068	     &0.07     &0.083	    &   0.03   &0.071	   &0.14        &0.034	  &0.12        &0.066  \\  
HD\,224679      & $-$0.19      &$-$       &0.09        &0.095        &$-$0.08      &0.030       &$-$0.06       &0.070    &$-$0.07     &0.025    &0.02      &0.066      &0.06     &0.074     &   0.01   &0.145    &0.04        &0.074    &0.02        &0.102  \\  
\hline                      
\end{tabular}
\label{tableB}
\end{table}
\end{landscape}

\begin{landscape}
\vspace*{\fill}
\renewcommand{\thetable}{B}
\setcounter{table}{0}
\begin{table}
\centering
\setlength{\tabcolsep}{0.5mm} 
\begin{tabular}{lrcrcrcrcrcrcrcrcrcrcrc}
\hline
\hline  
 Star Name   & [Ni/Fe]     &	error	   & [Cu/Fe]	  &  error        &   [Sr\,I/Fe]&	error    & [Sr\,II/Fe] &	error	&  [Y/Fe]		   &error	   &[Zr/Fe]	   &error	    &[Ba/Fe]   & error	   &[La/Fe]&	 error   &	[Ce/Fe]  &error	     &[Nd/Fe]	&error  &	[Eu/Fe]	     & error \\                 
 \hline                                                                                                                                                                                                                                                                                          
 HD\,2946    &0.03        &0.030 	   & $-$0.09	    & 0.035        & 0.18          & $-$     &  $-$           & $-$     & 0.14      &0.030 	  & 0.30  	  &$-$	    & 0.48     & 0.025      &    0.48     & $-$	 &   0.42	 	 &0.036	      &0.48			&$-$	    &  0.25      & $-$   \\                      
 HD\,3458    &$-$0.04     &0.020 	   & $-$0.23 	    & 0.010        & $-$0.06       & $-$     &  $-$           & $-$     & 0.05      &0.062	    & 0.12 	    &$-$	    & 0.38	   & 0.085      &    0.18     & $-$	 &   0.43	   &0.152	      &0.43			&$-$	    &  0.10      & $-$   \\                      
 HD\,4395    &0.00        &0.020 	   & $-$0.11 	    & 0.010        & 0.79          & $-$     &  0.70          & $-$     & 0.55      &0.096	    & 0.60  	  &$-$	    & 0.86	   & 0.045      &    0.60     & $-$	 &   0.67	   &0.110 	    &0.60 	  &$-$	    &  0.15      & $-$   \\                      
 HD\,11131   &$-$0.08     &0.020 	   & $-$0.26	    & 0.015        & 0.18          & $-$     &  0.17         & $-$     & 0.10 	   &0.045	    & 0.13 	    &$-$	    & 0.40	   & 0.000      &    0.16     & $-$	 &   0.28	   &0.025	      &0.22			&$-$	    &  0.16      & $-$   \\                      
 HD\,12484   &$-$0.12     &0.045	   & $-$0.23	    & 0.055        & 0.24          & $-$     &  0.16         & $-$     & 0.04      &0.088	    & 0.14 	    &$-$	    & 0.40     & 0.005      &    0.18	    & $-$  &   0.22 	 &0.020 	    &0.13			&$-$	    &  0.03 	   & $-$   \\                     
 HD\,16178   &0.17        &0.000     & $-$0.10      & 0.035        & 0.31          & $-$     &  0.30          & $-$     & 0.29      &0.094	    & 0.32 	    &$-$	    & 0.32	   & 0.075      &    0.20	    & $-$	 &   0.26 	 &0.074	      &0.44			&$-$	    &  $-$       & $-$   \\                
 HD\,18015   &$-$0.10     &0.030 	   & $-$0.18 	    & 0.010        & 0.02          & $-$     &  $-$ 0.02        & $-$     & 0.09      &0.010 	  & 0.13 	    &$-$	    & 0.31	   & 0.010      &    0.26 	  & $-$	 &   0.27	   &0.017	      &0.27			&$-$	    &  0.20      & $-$   \\                      
 HD\,18645   &$-$0.18     &0.030 	   & $-$0.21	    & 0.015        & 0.01          & $-$     &  0.00            & $-$     & $-$0.03   &0.134	    & 0.12 	    &$-$	    & 0.34	   & 0.040      &    0.02	    & $-$	 &   0.05	   &0.075	      &	$-$			&$-$	    &  $-$0.12	 & $-$   \\                      
 HD\,22233   &0.16        &0.025	   & $-$0.14      & 0.005        & 0.32          & $-$     &  0.1          & $-$     & 0.07      &0.074	    & 0.15 	    &$-$	    & 0.50	   & 0.070      &    0.28	    & $-$  &	 0.50    &0.167       &0.58			&$-$	    &  0.20  	   & $-$   \\                    
 HD\,38949   &$-$0.13     &0.090 	   & $-$0.22 	    & 0.020        & 0.13          & $-$     &  0.05         & $-$     & 0.09      &0.084	    & 0.27 	    &$-$	    & 0.34	   & 0.005      &    0.23 	  & $-$	 &   0.20  	 &0.000      	&0.17			&$-$	  	&  $-$       & $-$   \\                   
 HD\,45210   &$-$0.03     &0.000     & $-$0.20      & 0.005        & 0.18          & $-$     &  0.1          & $-$     & 0.09      &0.098	    & 0.03 	    &$-$	    & 0.40     & 0.020      &    0.21	    & $-$	 &   0.22 	 &0.079	      &0.23			&$-$	    &  0.12      & $-$   \\                    
 HD\,72440   &$-$0.07     &0.000     & $-$0.19      & 0.010        & 0.00          & $-$     &  0.00            & $-$     & 0.05      &0.005	    & 0.01 	    &$-$	    & 0.34	   & 0.015      &    0.15	    & $-$	 &   0.15	   &0.048	      &0.15			&$-$	    &  0.10      & $-$   \\                    
 HD\,103847  &$-$0.03     &0.025	   & $-$0.20      & 0.080        & 0.34          & $-$     &  $-$           & $-$     & 0.09      &0.063	    & 0.25 	    &$-$	    & 0.24 	   & 0.020      &    0.25	    & $-$	 &   0.29	   &0.054	      &0.25			&$-$	    &  0.22      & $-$   \\                    
 HD\,108189  &$-$0.14     &0.005	   & $-$0.23      & 0.005        & 0.01          & $-$     &  $-$ 0.03        & $-$     & $-$0.02   &0.005	    & 0.01      &$-$	    & 0.40	   & 0.005      &    0.08 	  & $-$	 &   0.10  	 &0.085	      &0.10 		&$-$	    &  0.10      & $-$   \\                    
 HD\,200491  &0.02        &0.075	   & $-$0.25      & 0.030        & $-$0.04       & $-$     &  $-$ 0.13        & $-$     & 0.07      &0.090 	  & 0.09 	    &$-$	    & 0.39 	   & 0.010      &    0.21	    & $-$	 &   0.29 	 &0.081	      &0.35			&$-$	    &  0.16      & $-$   \\                    
 HD\,205163  &0.08        &0.025	   & 0.04         & 0.000        & 0.06 	       & $-$     &  $-$           & $-$     & 0.05      &0.065	    & 0.15 	    &$-$	    & 0.35 	   & 0.010      &    0.28 	  & $-$	 &   0.25	   &0.074	      &0.28			&$-$	    &  0.30      & $-$   \\                    
 HD\,220122  &0.19        &0.090     & 0.05  	      & 0.050        & $-$0.14       & $-$     &  $-$           & $-$     & $-$0.04   &0.064	    & $-$0.01	  &$-$	    & 0.31  	 & 0.040      &    0.18 	  & $-$	 &   0.27		 &0.124	      &0.39			&$-$	    &  0.23      & $-$   \\                    
 HD\,224679  &$-$0.06     &0.010     & $-$0.11 	    & 0.040        & 0.10          & $-$     &  0.09         & $-$     & 0.09      &0.070 	  & $-$0.05	  &$-$	    & 0.32 	   & 0.035      &    0.12 	  & $-$	 &   0.19		 &0.045	      &0.04			&$-$	    &  0.05      & $-$   \\                    

  \hline                          
\end{tabular}
\end{table}
\end{landscape}

\end{appendix}
\end{document}